\documentclass[sigconf,natbib=true,anonymous=false]{acmart}

\AtBeginDocument{%
  }

\setcopyright{acmlicensed}
\copyrightyear{2018}
\acmYear{2018}
\acmDOI{XXXXXXX.XXXXXXX}
\acmConference[Conference acronym 'XX]{Make sure to enter the correct
  conference title from your rights confirmation emai}{June 03--05,
  2018}{Woodstock, NY}
\acmISBN{978-1-4503-XXXX-X/18/06}

\usepackage{multirow} 
\usepackage{pdflscape}
\usepackage[ruled,vlined]{algorithm2e}
\usepackage{enumitem}

\begin{document}

\title{AlignPxtr: Aligning Predicted Behavior Distributions for Bias-Free Video Recommendations}

\author{Chengzhi Lin}
\affiliation{%
  \institution{Kuaishou Technology}
  \city{Beijing}
  \country{China}
}
\email{1132559107@qq.com}

\author{Chuyuan Wang}
\affiliation{%
  \institution{Kuaishou Technology}
  \city{Beijing}
  \country{China}
}
\email{wangchuyuan@kuaishou.com}

\author{Annan Xie}
\affiliation{%
  \institution{Peking University}
  \city{BeiJing}
  \country{China}
}
\email{2301210470@stu.pku.edu.cn}

\author{Wuhong Wang}
\affiliation{%
  \institution{University of Science and Technology of China}
  \city{Hefei}
  \country{China}
}
\email{wangwuhong@mail.ustc.edu.cn}

\author{Ziye Zhang}
\affiliation{%
  \institution{Kuaishou Technology}
  \city{BeiJing}
  \country{China}
}
\email{zhangziye03@kuaishou.com}

\author{Canguang Ruan}
\affiliation{%
  \institution{Kuaishou Technology}
  \city{BeiJing}
  \country{China}
}
\email{ruancanguang@kuaishou.com}

\author{Yuancai Huang}
\affiliation{%
  \institution{Kuaishou Technology}
  \city{BeiJing}
  \country{China}
}
\email{huangyuancai@kuaishou.com}

\author{Yongqi Liu}
\affiliation{%
  \institution{Kuaishou Technology}
  \city{Beijing}
  \country{China}
}
\email{liuyongqi@kuaishou.com}

\renewcommand{\shortauthors}{Trovato et al.}

\begin{abstract}

In video recommendation systems, user behaviors such as watch time, likes, and follows are commonly used to infer user interest. However, these behaviors are influenced by various biases, including duration bias, demographic biases, and content category biases, which obscure true user preferences. In this paper, we hypothesize that biases and user interest are independent of each other. Based on this assumption, we propose a novel method that aligns predicted behavior distributions across different bias conditions using quantile mapping, theoretically guaranteeing zero mutual information between bias variables and the true user interest. By explicitly modeling the conditional distributions of user behaviors under different biases and mapping these behaviors to quantiles, we effectively decouple user interest from the confounding effects of various biases. Our approach uniquely handles both continuous signals (e.g., watch time) and discrete signals (e.g., likes, comments), while simultaneously addressing multiple bias dimensions. Additionally, we introduce a computationally efficient mean alignment alternative technique for practical real-time inference in large-scale systems.
We validate our method through online A/B testing on two major video platforms: Kuaishou Lite and Kuaishou. The results demonstrate significant improvements in user engagement and retention, with \textbf{cumulative lifts of 0.267\% and 0.115\% in active days, and 1.102\% and 0.131\% in average app usage time}, respectively. The results demonstrate that our approach consistently achieves significant improvements in long-term user retention and substantial gains in average app usage time across different platforms.
Our core code will be publised at https://github.com/justopit/CQE.

\end{abstract}

\ccsdesc[500]{Information systems~Recommender systems}

\keywords{Debiasing, User Behavior Alignment, Video Recommendations}



\maketitle
\section{Introduction}
Video recommendation systems have become an integral component of online platforms, helping users navigate vast content libraries by surfacing relevant videos based on their historical interactions. These systems play a critical role in shaping user experience, content discovery, and engagement metrics across streaming services, social media, and video-sharing platforms~\cite{video1, video2, video3, video4,micro_MARNET,micro_len,DCN,DCNV2,yang2024swat}. At their core, recommendation algorithms rely on observed user behaviors—such as watch time, likes, follows, and shares—to infer user preferences and predict interest in candidate content.

However, these observed behaviors are fundamentally confounded by various biases that obscure true user preferences. For instance, a user might watch longer videos simply due to their duration rather than genuine interest in the content. Similarly, demographic factors and content category preferences can systematically influence user behaviors in ways that do not reflect authentic interest. These confounding factors include, but are not limited to:

\begin{itemize}
    \item \textbf{Duration bias}: Users naturally spend more time on longer videos, regardless of their actual interest level. This creates a systematic preference for recommending longer content.
    \item \textbf{Demographic bias}: User behaviors vary across demographic groups due to cultural, socioeconomic, and accessibility factors, potentially leading to unfair recommendations across different user segments.
    \item \textbf{Content category bias}: Certain content categories inherently generate different behavior patterns—for example, educational videos may receive fewer likes than entertainment videos, even when users find them equally valuable. 
\end{itemize}

Other biases may include popularity bias, recency bias, presentation bias, and platform-specific interaction biases—all of which further complicate the task of inferring genuine user interest from observed behaviors. 

These biases create a fundamental challenge: \textbf{how can recommendation systems distinguish between behaviors driven by genuine user interest and those influenced by confounding biases?} Without addressing this challenge, recommendation algorithms risk optimizing for biased behavioral signals rather than true user preferences, leading to suboptimal user experiences, reduced engagement, and potential fairness concerns.

Existing approaches to debiasing in video recommendations have significant limitations. Most current methods focus narrowly on addressing a single bias dimension (usually video duration)~\cite{D2Q, D2Co, WTG} while ignoring other important biases. They also lack theoretical guarantees about the independence between debiased signals and bias factors. Moreover, existing techniques are generally designed for debiasing a single behavioral signal (e.g. watch time) rather than integrating multiple types of user feedback.

In this paper, we introduce a novel framework for bias-free video recommendations based on a key insight: \textbf{user interest and biases are inherently independent.} This independence assumption forms the theoretical foundation of our approach. We propose a causal framework where observed user behaviors ($S$) are influenced by both user interest ($Z$) and various biases ($Y$), mediated through a latent behavior distribution ($X$). By explicitly modeling this causal structure, we develop a principled method to decouple true user interest from confounding biases.

The core technical contribution of our work is a \textbf{distribution alignment} technique that transforms biased behavioral signals into bias-independent user interest scores. We prove that by aligning the conditional distributions of predicted behaviors across different bias conditions through quantile mapping, we can achieve zero mutual information between the bias variables and the user interest score. This theoretical result guarantees that our method produces interest scores that are truly independent of the underlying biases. 

Our approach offers several key advantages over existing methods:

\begin{enumerate}
\item \textbf{Theoretical guarantees}: We provide formal proof that our quantile mapping approach achieves independence between bias factors and extracted user interest, unlike heuristic approaches that lack such guarantees.

\item \textbf{Multi-bias handling}: Our framework accommodates multiple bias dimensions simultaneously, rather than focusing solely on duration bias or any single confounding factor.

\item \textbf{Multi-signal integration}: Unlike existing approaches that typically only debias a single continuous signal like watch time, our method works effectively for integrating both continuous behavioral signals (e.g., watch time) and discrete/binary signals (e.g., likes, comments, follows) into a unified interest representation.

\item \textbf{Computational efficiency}: We introduce a mean alignment alternative that maintains most benefits of quantile mapping while being substantially more practical for real-time inference in large-scale online systems.
\end{enumerate}

Through extensive online experiments on Kuaishou Lite and Kuaishou , we observe substantial cumulative gains of 0.267\% and 0.115\% in user active days, along with 1.102\% and 0.131\% improvements in average app usage time. These results validate the effectiveness of our approach in enhancing long-term user engagement and retention in real-world large-scale video recommendation scenarios.

In summary, the key contributions of our work are:

\begin{itemize}
    \item We formalize the hypothesis that biases and user interest are independent, and prove that quantile mapping of predicted behavior distributions leads to zero mutual information between biases and  the user interest score.
    
    \item We propose a unified framework for decoupling multiple bias factors from various types of user behaviors using conditional quantile mapping, effectively isolating user interest from biases. 
    
    \item We present an alternative  practical solution for efficient real-time inference in large-scale online systems through mean alignment of candidate behavior distributions.
    
    \item We demonstrate the effectiveness of our approach through online experiments on real-world video recommendation platforms with hundreds of millions of users.
\end{itemize}

\section{Method}

\subsection{Problem Formulation}
In video recommendation systems, user behaviors (e.g., watch time, likes, follows) are influenced by both genuine user interest and various biases, such as video duration, demographic characteristics, and content popularity. This entanglement presents a fundamental challenge: \textbf{accurately modeling a user's true interest while minimizing the influence of confounding biases}. Our objective is to decouple true user interest from biases embedded in observed behaviors, ensuring that recommendations reflect authentic user preferences rather than artifacts of the recommendation system itself.

\subsection{Causal Framework}
\begin{figure}[t]
    \centering
    \includegraphics[width=0.8\columnwidth]{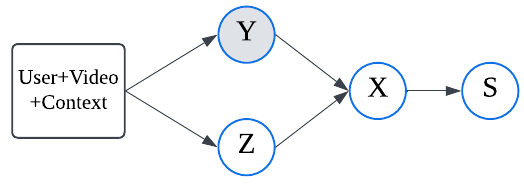}
    \caption{Causal graph illustrating the relationships between user+video+context, bias factors ($Y$), user interest ($Z$), latent behavior distribution ($X$), and observed user behavior ($S$). The bias factors ($Y$) independently influence user interest ($Z$). User interest ($Z$) and bias factors ($Y$) jointly affect the latent behavior distribution ($X$), which in turn determines the observed user behavior ($S$).}
    \label{fig:causal_graph}
\end{figure}

We propose a causal framework that explicitly models the relationships between key variables in the recommendation process, as illustrated in Figure \ref{fig:causal_graph}:

\begin{itemize}
    \item $Z$: True user interest - the unbiased latent preferences we aim to uncover
    \item $Y$: Bias-related factors (video duration, demographics, content popularity)  
    \item $X$: Latent behavior distribution influenced jointly by $Z$ and $Y$
    \item $S$: Observed user behavior (watch time, likes, follows, etc.)
\end{itemize}

The causal relationships among these variables follow this structure:
\begin{enumerate}
    \item $Z$ and $Y$ independently exist as root causes in the causal graph
    \item $Z$ and $Y$ jointly influence $X$, which represents a latent behavior distribution 
    \item $X$ determines the observed behavior $S$
\end{enumerate}

Our primary objective is to recover $Z$ (true user interest) by decoupling it from the effects of $Y$ (biases). This requires \textbf{minimizing the mutual information} $I(Y;Z)$ to ensure independence between bias factors and user interest. Following the causal structure, $X$ can be recovered from $Z$ and $Y$ since $X = f(Z,Y)$ for some function $f$, allowing us to predict observed behaviors $S$ through $X$ even after extracting the bias-free interest $Z$.

\subsection{Bias Decoupling via Distribution Alignment}

\subsubsection{Theoretical Foundation} 
The core insight of our approach is that the conditional distribution of $Z$ given any value of $Y$ should be identical across all possible values of $Y$, ensuring zero mutual information between $Z$ and $Y$. This condition guarantees that $Z$ (user interest) and $Y$ (biases) are truly independent. Formally, the mutual information between $Y$ and $Z$ is defined as:

\begin{equation}
    I(Y;Z) = \int_y \int_z p(y,z) \log \left( \frac{p(y,z)}{p(y)p(z)} \right) dz dy
    \label{eq:mutual_info}
\end{equation}

When the conditional distribution $p(z|y)$ is invariant across all values of $y$, then $p(y,z) = p(y)p(z)$, and the mutual information becomes zero:

\begin{equation}
    I(Y;Z) = \int_y \int_z p(y)p(z) \log \left( \frac{p(y)p(z)}{p(y)p(z)} \right) dz dy = 0
    \label{eq:zero_mutual_info} 
\end{equation}

Therefore, our goal is to find a transformation that maps the latent behavior variable $X$ to a bias-independent user interest $Z$ such that the conditional distribution of $Z$ given $Y$ remains invariant to $Y$.

\subsubsection{Conditional Quantile Mapping}
Based on this theoretical foundation, we propose using \textbf{conditional quantile mapping} to transform $X$ into $Z$. For any observation pair $(x,y)$, we first estimate the conditional cumulative distribution function (CDF) of $X$ given $Y=y$, denoted as $F_{X|Y=y}$. We then compute the quantile $\tau$ of $x$ under this conditional distribution:

\begin{equation}
    \tau = F_{X|Y=y}(x)
    \label{eq:quantile}
\end{equation}

We directly use $\tau$ as the bias-decoupled user interest $Z$:

\begin{equation}
    z = \tau = F_{X|Y=y}(x)  
    \label{eq:z_tau}
\end{equation}

This transformation ensures that $Z$ follows a uniform distribution on $[0,1]$, regardless of the value of $Y$, effectively decoupling user interest from bias. The key property exploited here is that the random variable $F_{X|Y=y}(X) \sim \text{Uniform}(0,1)$ for any continuous distribution, which guarantees that $Z$ is independent of $Y$ by construction.

\begin{proposition}
Let $X$ be a continuous random variable with conditional CDF $F_{X|Y=y}$ given $Y=y$. Define $Z = F_{X|Y=y}(X)$. Then $Z \perp Y$ (i.e., $Z$ is independent of $Y$). 
\end{proposition}

\begin{proof}
For any $z \in [0,1]$ and any value $y$ of $Y$:
\begin{align}
    P(Z \leq z | Y=y) &= P(F_{X|Y=y}(X) \leq z | Y=y) \nonumber \\
    &= P(X \leq F_{X|Y=y}^{-1}(z) | Y=y) \nonumber \\
    &= z \nonumber
\end{align}
Since the conditional distribution of $Z$ given $Y=y$ is uniform on $[0,1]$ for all values of $y$, $Z$ is independent of $Y$.
\end{proof}

In practice, we can estimate the conditional distribution $F_{X|Y}$ using either non-parametric methods (e.g., kernel density estimation, empirical distribution functions, quantile regression) or parametric methods (e.g., assuming a specific distribution family and estimating how its parameters depend on $Y$).

\subsubsection{ From Uniform Distribution to Domain-Specific Distribution}

The uniform distribution $Z \in [0, 1]$ obtained through conditional quantile mapping provides a representation that is independent of bias factors $Y$. However, in practical applications, we may need to transform this uniform distribution into a domain-specific distribution that better aligns with business requirements.

Using the principle of Inverse Transform Sampling~\cite{Wikipedia_InverseTransformSampling}, we can convert the uniform distribution $Z$ into any desired target distribution $G$:

$$Z' = F^{-1}_G(Z)$$

where $F^{-1}_G$ is the inverse cumulative distribution function of the target distribution $G$. Since $Z$ is independent of $Y$, the transformed $Z'$ maintains this independence while exhibiting distributional properties more suitable for the application domain.

The choice of target distribution $G$ can be based on:
1. Domain knowledge (e.g., expected distribution of user interest scores)
2. Empirically observed distribution of true user interests in historical data
3. Optimization objectives (e.g., adjusting the distribution to maximize ranking performance)

This two-step transformation approach (first removing bias to obtain a uniform distribution, then transforming to a domain-specific distribution) preserves our theoretical guarantees while enhancing the method's flexibility and utility in practical applications.

\subsubsection{Mean Alignment as a Practical Alternative}
In real-world applications where bias variables $Y$ may vary for each user and change over time, estimating the full conditional distribution $F_{X|Y}$ for each user-time combination becomes computationally challenging due to limited samples and inference time constraints. As a more efficient alternative, we propose aligning the means of the conditional distributions:

\begin{equation}
    z = x - \mathbb{E}[X|Y=y]
    \label{eq:mean_align}
\end{equation}

where $x$ is the realized value of the latent behavior variable $X$, $y$ is the realized value of the bias variable $Y$, and $\mathbb{E}[X|Y=y]$ is the conditional expectation of $X$ given $Y=y$. This transformation centers the conditional distributions to a common mean, effectively removing the first-order influence of bias $Y$ on the  expected behavior $X$.  

While this approach does not fully align the entire conditional distributions (and thus does not guarantee complete independence between $Z$ and $Y$), it serves as a computationally efficient alternative when:
\begin{itemize}
    \item The bias effects are primarily reflected in the means
    \item Real-time inference latency is a critical constraint
\end{itemize}

Empirically, we find that this mean alignment approach provides a favorable trade-off between computational efficiency and bias mitigation in large-scale recommendation systems.

\begin{table*}[h]
\centering
\caption{Experiment results on Kuaishou Lite. Each row represents the lift of the corresponding method compared to the combination of all previous methods. A boldface means a statistically significant result  (p-value $<$ 5\%).}
\label{tab:kuaishou_lite_results}
\begin{tabular}{lcccc}
\toprule
\textbf{Method} & \textbf{Active Days} & \textbf{App Time} & \textbf{Watchtime} & \textbf{Views} \\
\midrule
$\text{AlignPxtr}_{\text{mean, }Y_1}$ & \textbf{+0.077\%} & \textbf{+0.387\%} & \textbf{+1.053\%} & \textbf{+0.518\%} \\
$+ \text{AlignPxtr}_{\text{mean, }Y_2}$ & \textbf{+0.078\%} & \textbf{+0.622\%} & \textbf{+2.107\%} & \textbf{-1.669\%} \\
$+ \text{AlignPxtr}_{\text{mean, }Y_3}$ & \textbf{+0.061\%} & \textbf{+0.093\%} & \textbf{+0.177\%} & \textbf{-0.151\%} \\
$+ \text{AlignPxtr}_{\text{quantile, }Y_4}$ & \textbf{+0.051\%} & +0.015\% & -0.005\% & \textbf{+0.258\%} \\
\bottomrule
\end{tabular}
\end{table*}

\subsection{Model Training and Inference}

\subsubsection{Training Procedure}
Our training procedure consists of two main steps:

\paragraph{Step 1: Behavior Modeling} Estimate the latent behavior distribution $X$ by fitting a model to the observed behavior data $S$:
\begin{itemize}
    \item For continuous behaviors (e.g., watch time): Minimize mean squared error loss
    \item For binary behaviors (e.g., likes, follows): Minimize binary cross-entropy loss
\end{itemize}

The model architecture can be any standard recommendation model (e.g., neural networks, matrix factorization) that takes user and item features as input and predicts user behaviors.

\paragraph{Step 2: Bias Modeling} Learn the conditional statistics of $X$ given $Y$:
\begin{itemize}
    \item For quantile mapping: Estimate $F_{X|Y}$ using non-parametric or parametric methods
    \item For mean alignment: Calculate $\mathbb{E}[X|Y]$ for different bias conditions 
\end{itemize}

To handle multiple bias dimensions simultaneously, we can either: 
\begin{enumerate}
    \item Discretize the bias space and estimate separate conditional distributions for each combination
    \item Use parametric models to capture how distribution parameters depend on bias variables 
    \item Employ non-parametric regression techniques to estimate conditional statistics, such as CQE~\cite{CQE}
\end{enumerate}

\subsubsection{Inference}
At inference time, given a user-item pair $(u,v)$ with bias vector $y$:
\begin{enumerate}
\item Predict the latent behavior variable $x$ using the pre-trained behavior model
\item Apply the appropriate transformation to obtain the bias-decoupled user interest $z$:
\begin{itemize}
    \item Quantile mapping: $z = F_{X|Y=y}(x)$
    \item Mean alignment: $z = x - \mathbb{E}[X|Y=y]$
\end{itemize}
\end{enumerate}

For multi-dimensional behaviors (watch time, likes, follows, comments), we compute the decoupled interest score for each dimension separately and combine them using importance weights:

\begin{equation}
    z_{\text{final}} = \sum_{i\in\text{behaviors}} w_i \cdot z_i
    \label{eq:multi_dim}
\end{equation}

where $w_i$ are importance weights for each behavior dimension, which can be determined through offline evaluation or online A/B testing.

\section{Experiments}
\subsection{Online Experiments Setup}
To validate the effectiveness of our proposed AlignPxtr framework, we conducted extensive online experiments on two major video platforms: Kuaishou Lite and Kuaishou. Each platform has hundreds of millions of daily active users, providing a robust testbed for evaluating the real-world impact of our debiasing approach.

For each platform, we randomly split users into control and treatment groups, with each group receiving over 10\% of the total traffic. The experiments ran over two weeks, ensuring sufficient data collection for reliable analysis. Our experimental domain focused on optimizing the main page of each app, as improvements in the main page experience can drive overall gains in key engagement metrics such as average app usage time and active days per user.

\subsection{Evaluation Metrics}
We evaluated the performance of our method using the following metrics:
\begin{itemize}
    \item \textbf{Active Days (LT7)}: The number of active days per user over a 7-day window, capturing user stickiness and retention.
    \item \textbf{Average App Usage Time}: The mean time spent by users in the app, indicating overall engagement.
    \item \textbf{Page Watchtime}: The total watch time on the main page, directly reflecting the quality of recommendations.
    \item \textbf{Video Views}: The number of video plays on the main page, measuring user interaction with recommended content.
\end{itemize}

\subsection{Results on Kuaishou Lite}
We conducted a series of experiments on Kuaishou Lite to evaluate different variants of AlignPxtr. Each variant builds upon the previous ones, incrementally incorporating additional bias factors and alignment techniques to provide a comprehensive evaluation of our approach.

As shown in Table~\ref{tab:kuaishou_lite_results}, we start with AlignPxtr$_{\text{mean}, Y_1}$, which applies mean alignment to the first bias variable $Y_1$. This initial implementation achieves significant improvements over the baseline across all metrics, with $+0.077\%$ increase in active days, $+0.387\%$ in app usage time, $+1.053\%$ in watch time, and $+0.518\%$ in video views.

Building upon this foundation, we introduced AlignPxtr$_{\text{mean}, Y_2}$, which incorporates an additional bias factor $Y_2$ using the mean alignment technique. This addition yields further improvements: $+0.078\%$ in active days, $+0.622\%$ in app usage time, and $+2.107\%$ in page watch time. We observe a decrease of $-1.669\%$ in video views, but this trade-off is considered favorable as the gains in long-term user retention and overall engagement metrics outweigh this reduction.

Next, we implemented AlignPxtr$_{\text{mean}, Y_3}$, which addresses a third bias dimension. This variant delivers incremental lifts of $+0.061\%$ in active days, $+0.093\%$ in app usage time, and $+0.177\%$ in watch time, with a slight decrease of $-0.151\%$ in video views.

\begin{table*}[h]
\centering
\caption{Experiment results on Kuaishou.  A boldface means a statistically significant result  (p-value $<$ 5\%).}
\label{tab:kuaishou_results}
\begin{tabular}{lcccc}
\toprule
\textbf{Method} & \textbf{Active Days} & \textbf{App Time} & \textbf{Watchtime} & \textbf{Views} \\
\midrule

$\text{AlignPxtr}_{\text{quantile, }Y_4}$ & \textbf{+0.041\%} & \textbf{+0.061\%}  & \textbf{+0.103\%} & \textbf{+0.241\%} \\

$\text{+AlignPxtr}_{\text{mean, }Y_1}$ & \textbf{+0.074\%} & \textbf{+0.070\%} & \textbf{+0.148\%} & \textbf{+0.178\%} \\

\bottomrule
\end{tabular}
\end{table*}

Finally, we evaluated AlignPxtr$_{\text{quantile}, Y_4}$, which employs the more theoretically robust quantile mapping technique on bias variable $Y_4$. This approach provides additional gains of $+0.051\%$ in active days, $+0.015\%$ in app usage time, and $+0.258\%$ in video views, with a negligible impact of $-0.005\%$ on watch time.

In total, the cumulative improvements across all AlignPxtr variants amount to $+0.267\%$ in active days, $+1.102\%$ in app usage time, $+3.327\%$ in watch time. These results demonstrate the effectiveness of our progressive debiasing approach, with each additional component contributing meaningful improvements to key engagement metrics.

\subsection{Results on Kuaishou}
We further validated our approach on the main Kuaishou app to demonstrate the generalizability of AlignPxtr across different platforms. The experimental results are presented in Table~\ref{tab:kuaishou_results}.

Our first implementation on Kuaishou employed AlignPxtr$_{\text{quantile}, Y_4}$, applying quantile mapping to bias variable $Y_4$. This variant achieved statistically significant improvements across all metrics: $+0.041\%$ in active days, $+0.061\%$ in app usage time, $+0.103\%$ in watch time, and $+0.241\%$ in video views.

Building upon this foundation, we introduced AlignPxtr$_{\text{mean}, Y_1}$, which addresses an additional bias dimension using mean alignment. This enhancement provided further gains of $+0.074\%$ in active days, $+0.070\%$ in app usage time, $+0.148\%$ in watch time, and $+0.178\%$ in video views.

The cumulative improvements from both variants amount to $+0.115\%$ in active days, $+0.131\%$ in app usage time, $+0.251\%$ in watch time, and $+0.419\%$ in video views. These consistent gains across all key metrics validate the effectiveness of our AlignPxtr framework in real-world large-scale video recommendation scenarios.

Notably, the results on both Kuaishou Lite and Kuaishou demonstrate that our approach consistently achieves significant improvements in long-term user retention and substantial gains in engagement metrics, confirming the practical value of distribution alignment for debiasing video recommendations.

\subsection{Discussion}
The experimental results on Kuaishou Lite and Kuaishou provide strong evidence supporting the efficacy of our distribution alignment approach for debiasing video recommendations. By explicitly modeling the conditional distributions of user behaviors under different biases and aligning these distributions through quantile mapping or mean matching, AlignPxtr effectively decouples user interest from the confounding effects of various biases.

The consistent gains observed in key metrics such as active days  highlight the practical value of our debiasing framework in enhancing user experience and driving long-term engagement. These improvements are especially noteworthy given the large scale of the platforms, where even small percentage lifts translate to substantial absolute gains.

The results also demonstrate the flexibility of AlignPxtr in handling different bias variables and alignment techniques. By incorporating multiple bias factors and comparing mean alignment with quantile mapping, we showcase the adaptability of our framework to various real-world scenarios and the potential for further optimization based on domain-specific considerations.

Overall, the successful online experiments on Kuaishou Lite and Kuaishou validate the theoretical foundation and practical effectiveness of AlignPxtr, establishing it as a promising approach for mitigating biases and improving recommendation quality in large-scale video platforms.

\section{Related Work}

\subsection{Video Recommendation Systems}

Video recommendation systems play a crucial role in organizing and delivering massive video content to users, particularly with the surge of short video platforms such as TikTok and YouTube Shorts. Existing systems are commonly categorized into content-based, collaborative filtering, hybrid, and group-based methods~\cite{lubos2023overview}, each aiming to ensure personalized and efficient content consumption. However, challenges such as cold start, scalability, and inherent bias remain prevalent. In the domain of short videos, modeling temporal dynamics and user interests is especially important. Approaches like temporal hierarchical attention networks have been developed to capture both category- and item-level patterns in micro-video recommendation~\cite{chen2018temporal}. Furthermore, multi-modal representation has gained attention due to its ability to integrate visual, audio, and textual information for a more holistic understanding of video content, which also alleviates cold start issues~\cite{pingali2022towards, raj2023multi}. Recent work has explored graph-based methods, such as Graph Convolutional Networks and Graph Attention Networks, to enrich user and item embeddings by incorporating structural and latent features~\cite{wei2019mmgcn, ting2022micro, wang2021dualgnn}.
\subsection{Causal Inference in Recommendation}
Causal inference techniques have been increasingly applied to recommendation systems to enhance robustness, fairness, and generalization beyond conventional accuracy metrics. These methods address issues such as data bias, noise, and distribution shift~\cite{gao2024causal}. Backdoor adjustment, counterfactual inference, and deconfounding embeddings are representative approaches to mitigate popularity and exposure biases~\cite{zhang2021causal,wei2021model,schnabel2016recommendations}. Causal embeddings~\cite{bonner2018causal} and dynamic causal collaborative filtering~\cite{xu2022dynamic} have been particularly effective in capturing latent confounders and improving performance under sparse and noisy conditions. Causal methods also offer enhanced interpretability and robustness for short video platforms, where user interactions are often brief and behavior patterns more volatile. Techniques like causal preference learning~\cite{he2022causpref} and confounding-aware modeling~\cite{he2023addressing} further improve out-of-distribution generalization.

\subsection{Debiasing Techniques in Recommendation}

Debiasing is vital for building fair and balanced recommendation systems. A wide range of biases—including selection, exposure, and position biases—has been identified and categorized, along with corresponding debiasing strategies~\cite{chen2023bias}. Propensity score weighting, causal intervention, and uniform data strategies have been widely adopted to address these biases~\cite{wang2021deconfounded,yang2025multi}. MACR~\cite{wei2021model} eliminates popularity bias through multi-task learning, while CVRDD ~\cite{tang2023counterfactual} employs multi-task learning with learnable parameters to remove duration bias. D2Q ~\cite{zhan2022deconfounding} addresses the confounding effects of duration on item recommendations through backdoor adjustment techniques, and DVR~\cite{zheng2022dvr} introduces a new metric to achieve duration-unbiased recommendations. CWM ~\cite{zhao2024counteracting}and D²Co ~\cite{zhao2023uncovering}analyze user viewing behaviors in detail, focusing on duration bias from the perspectives of counterfactual watch time and noisy viewing patterns, respectively, to better reveal users’ true interests. In contrast to these methods, which target specific biases, our approach provides a unified framework that models all bias features simultaneously, achieving a decoupling of bias from true user interest.

\section{Conclusion}
In this paper, we presented AlignPxtr, a novel framework for debiasing video recommendations by aligning predicted user behavior distributions across different bias conditions. Our approach is built upon the key insight that user interest and biases are inherently independent, and by explicitly modeling the causal relationships between these factors, we can effectively decouple true user preferences from confounding biases.

AlignPxtr achieves theoretically guaranteed bias independence while maintaining computational efficiency for real-time inference in large-scale systems. The framework seamlessly integrates multiple bias dimensions and behavior types, providing a unified solution for mitigating the impact of various biases on video recommendations.
We validated the effectiveness of AlignPxtr through extensive online A/B testing on Kuaishou Lite and Kuaishou, demonstrating significant improvements in user engagement and retention, with cumulative lifts of 0.267\% and 0.115\% in active days, and 1.102\% and 0.131\% in average app usage time, respectively. These gains highlight the practical value of our debiasing approach in enhancing user experience and driving long-term retention.

By providing a principled and effective framework for decoupling user interest from confounding biases, AlignPxtr represents a significant step forward in the quest for unbiased video recommendations. As the demand for personalized content continues to grow, we believe that AlignPxtr will play an increasingly critical role in shaping the future of video recommendation systems.

\bibliographystyle{ACM-Reference-Format}
\bibliography{sample-base}

\appendix

\end{document}